\documentstyle[aps,prl,preprint,floats,epsfig]{revtex}

\begin{document}

\preprint{\tighten\vbox{\hbox{\hfil CLNS 00/1671}
                        \hbox{\hfil CLEO 00-8}
}}

\title{Study of exclusive two-body {\boldmath $B^0$} meson decays to charmonium}

\author{CLEO Collaboration}
\date{\today}

\maketitle
\tighten

\begin{abstract} 
We present a study of three  $B^0$ decay modes useful for 
time-dependent $CP$ asymmetry measurements. 
From  a sample of   $9.7\times10^6$ $B \overline B$ meson pairs collected with the CLEO detector,   we have reconstructed $B^0 \to J/\psi \, K^{0}_S$,  $B^0 \to \chi_{c1} \, K^{0}_S$, and $B^0 \to J/\psi \, \pi^{0}$ decays.
The latter two decay modes have been observed for the first time. 
We describe  a $K^{0}_S \to \pi^0 \pi^0$ detection technique  and its application to the reconstruction   of the decay $B^0 \to J/\psi \, K^{0}_S$.
Combining the results obtained using  $K^0_S \to \pi^+ \pi^-$ and $K^0_S \to \pi^0 \pi^0$ decays, we determine  ${\cal B}(B^0 \to J/\psi \, K^0)  = (9.5 \pm 0.8 \pm 0.6)\times10^{-4}$, where the first uncertainty 
is statistical and the second one is systematic.
 We also obtain ${\cal B}(B^0 \to \chi_{c1}\, K^0)  = 
(3.9^{+1.9}_{-1.3}\pm 0.4)\times10^{-4}$ and ${\cal B}(B^0 \to
J/\psi  \,   \pi^0)  =  (2.5^{+1.1}_{-0.9}\pm 0.2)\times  10^{-5}$. 
\end{abstract}
\newpage

{
\renewcommand{\thefootnote}{\fnsymbol{footnote}}

\begin{center}
P.~Avery,$^{1}$ C.~Prescott,$^{1}$ A.~I.~Rubiera,$^{1}$
J.~Yelton,$^{1}$ J.~Zheng,$^{1}$
G.~Brandenburg,$^{2}$ A.~Ershov,$^{2}$ Y.~S.~Gao,$^{2}$
D.~Y.-J.~Kim,$^{2}$ R.~Wilson,$^{2}$
% ...
\end{center}
 
\small
\begin{center}
$^{1}${University of Florida, Gainesville, Florida 32611}\\
$^{2}${Harvard University, Cambridge, Massachusetts 02138}\\
% ...
\end{center}
 
\setcounter{footnote}{0}
}
\newpage

$CP$ violation arises  naturally in the Standard Model with three
quark generations~\cite{CKM}; however, it  still remains one of the
least experimentally constrained sectors of the Standard Model.
Measurements of time-dependent rate asymmetries in the decays of
neutral $B$ mesons will provide an important test of the Standard
Model mechanism for $CP$ violation~\cite{Nir:1999}.

In this Article, we present a study of  $B^0 \to J/\psi \, K^{0}_S$,
$B^0 \to \chi_{c1} \, K^{0}_S$, and $B^0 \to J/\psi \, \pi^{0}$
decays.   The latter two decay modes have been observed for the first
time.  We describe  a $K^{0}_S \to \pi^0 \pi^0$ detection technique
and its application to the reconstruction   of the decay $B^0 \to
J/\psi \, K^{0}_S$.

The measurement of the $CP$ asymmetry in $B^0(\overline {B}^0)\to
 J/\psi \, K^0_S$ decays probes  the relative weak phase between the
 $B^0 - \overline {B}^0$ mixing amplitude and the $b \to c \bar c s$
 decay amplitude~\cite{sin2b}.  In the Standard Model this measurement
 determines   $\sin 2\beta$, where   $\beta \equiv {\rm Arg} \left (-
 V_{cd}V^*_{cb}/V_{td}V^*_{tb} \right)$.   A measurement of $\sin
 2\beta$ with $B^0(\overline {B}^0)\to \chi_{c1} \, K^0_S$ decays  is
 as theoretically clean as one with $B^0(\overline {B}^0)\to J/\psi \,
 K^0_S$.

For the purposes of $CP$ violation measurements, the $B^0\to J/\psi \,
\pi^0$  decay is similar to  $B^0\to D^+ \, D^-$: both decays are
governed by the $b \to c \bar c d$ quark transition, and  both final
states are $CP$ eigenstates of the same $CP$ sign. A recent search for
the $B^0\to D^+ \, D^-$ decay at CLEO   established  an upper limit on
${\cal B}(B^0\to D^+ \, D^-)$~\cite{CLEO-Dplus-Dminus}.  If the
penguin  ($b \to d  c \bar c$) amplitude is negligible compared to the
tree ($b \to c \bar c d$) amplitude, then the  measurement the $CP$
asymmetry in $B^0(\overline {B}^0)\to J/\psi \,  \pi^0$   decays
allows a  theoretically clean extraction of $\sin 2\beta$. The
asymmetries measured with $J/\psi\,  K^0_S$ and $J/\psi \, \pi^0$
final states should have exactly the same absolute values but
opposite signs, thus  providing  a useful check for  charge-correlated
systematic bias in $B$-flavor tagging.  If the ratio of penguin to
tree amplitudes is not  too small~\cite{Ciuchini}, then comparison of
the measured asymmetries in $J/\psi \,  K^0_S$ and $J/\psi \, \pi^0$
modes may allow a  resolution of one of the two   discrete ambiguities
($\beta \to \beta+\pi$) remaining after  a $\sin2\beta$
measurement~\cite{Grossman:1997gd}.

The data  were collected at   the Cornell Electron Storage Ring
(CESR) with    two   configurations    of     the   CLEO detector
called CLEO~II~\cite{Kubota:1992ww}  and
CLEO~II.V~\cite{Hill:1998ea}.  The components of the CLEO detector
most relevant to this analysis are the charged particle tracking
system, the CsI electromagnetic calorimeter, and the muon chambers.
In CLEO~II the momenta of charged particles are measured in a tracking
system consisting of a 6-layer straw tube chamber,  a 10-layer
precision drift chamber, and a 51-layer main drift chamber, all
operating inside a 1.5 T  solenoidal magnet. The main drift chamber
also provides a measurement of the  specific ionization, $dE/dx$, used
for particle identification.  For  CLEO~II.V, the straw tube  chamber
was replaced  with a  3-layer silicon vertex detector, and the gas in
the main drift chamber was changed from an argon-ethane to a
helium-propane mixture. The muon chambers  consist of proportional
counters placed at increasing depth in  the steel absorber.  We use
9.2~$\rm fb^{-1}$ of $e^+e^-$ data taken at the $\Upsilon(4S)$
resonance  and 4.6~$\rm fb^{-1}$ taken  60~MeV below the
$\Upsilon(4S)$ resonance.  Two thirds of the data  were collected with
the CLEO~II.V detector.  The simulated  event samples used in this
analysis were generated with a GEANT-based~\cite{GEANT} simulation of
the  CLEO detector response and  were processed in a similar manner as
the  data.  

We reconstruct  both  $J/\psi \to e^+ e^-$  and  $J/\psi \to \mu^+
\mu^-$  decays and use identical $J/\psi$ selection criteria for all
measurements described in this Article.  Electron candidates are
identified based on the ratio of the  track momentum to the associated
shower energy in the CsI calorimeter and on  the $dE/dx$  measurement.
The internal bremsstrahlung  in the  $J/\psi  \to  e^+ e^-$ decay  as
well  as the bremsstrahlung in the detector  material produces a long
radiative tail in the  $ e^+ e^-$   invariant mass distribution  and
impedes efficient $J/\psi  \to   e^+  e^-$  detection.   We  recover
some of the  bremsstrahlung photons by selecting the photon  shower
with the smallest opening angle with respect to the  direction of the
$e^\pm$ track evaluated at the interaction point, and then   requiring
this opening angle to be smaller than $5^\circ$. We therefore refer to
the $e^+ (\gamma) e^- (\gamma)$ invariant mass when we describe the
$J/\psi \to e^+ e^-$ reconstruction.  For the $J/\psi \to \mu^+ \mu^-$
reconstruction, one of the muon candidates is  required to penetrate
the steel absorber to a depth greater than 3  nuclear interaction
lengths.  We relax the absorber penetration  requirement for the
second muon candidate  if it is not expected to reach a muon chamber
either because its energy is too low or because it does not point to
a region of the  detector  covered by the muon  chambers.  For these
muon candidates  we require the ionization  signature in the  CsI
calorimeter to be  consistent with that of a muon.

\begin{figure}[htb]
\centering \epsfxsize=165mm \epsfbox{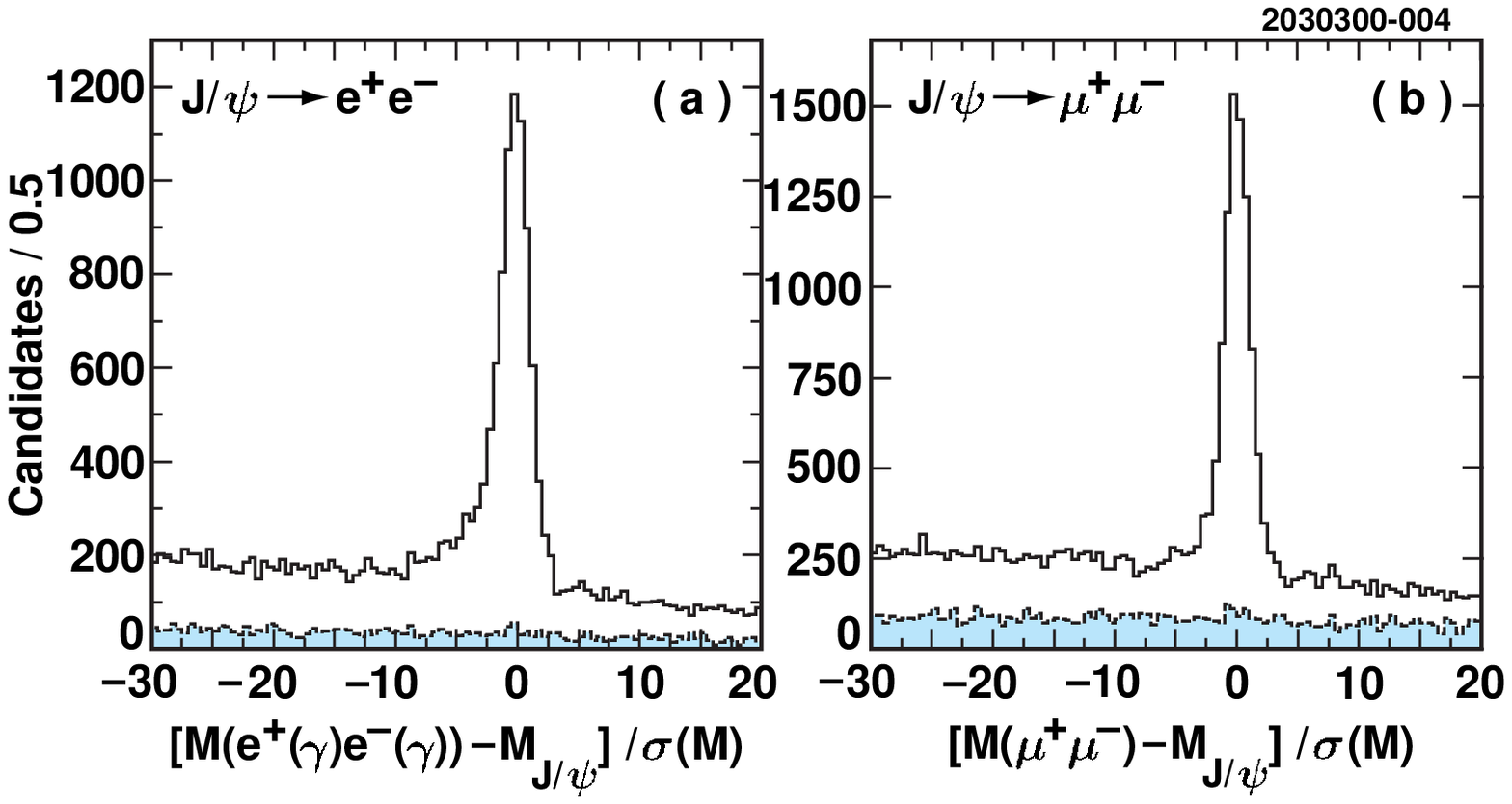}
\caption{Normalized invariant mass of the (a) $J/\psi \to e^+ e^-$ and
(b) $J/\psi  \to \mu^+  \mu^-$ candidates  in  data. The momentum of
the  $J/\psi$ candidates is required to be less than 2~GeV/$c$, which
is slightly above  the maximal $J/\psi$ momentum in  $B \to J/\psi \,
\pi$ decays.    The  shaded  histogram  represents the
luminosity-scaled  data taken  60~MeV below the $\Upsilon(4S)$
showing  the level of background from non-$B \overline B$ events.  } 
\label{fig:data_ee_mumu_on_off}
\end{figure}

We extensively use normalized variables, taking  advantage of
well-understood track and photon-shower four-momentum     covariance
matrices to calculate the expected resolution for each
combination. The use of normalized variables allows uniform candidate
selection criteria to be applied to the data collected with the
CLEO~II and CLEO~II.V detector configurations.  For example, the
normalized  $J/\psi \to \mu^+ \mu^-$ mass is defined as  $[M(\mu^+
\mu^-)-M_{J/\psi}]/\sigma(M)$, where $M_{J/\psi}$ is the world average
value of the $J/\psi$ mass~\cite{PDG} and $\sigma(M)$ is the
calculated mass resolution for that  particular  $\mu^+ \mu^-$
combination. The average $\ell^+ \ell^-$ invariant mass resolution  is
12~MeV$/c^2$.  The normalized mass distributions for the $J/\psi \to
\ell^+ \ell^-$     candidates   are         shown     in
Fig.~\ref{fig:data_ee_mumu_on_off}.   We   require     the normalized
mass  to be from $-10$  to $+3$  for the $J/\psi  \to  e^+  e^-$ and
from $-4$  to $+3$ for the  $J/\psi \to  \mu^+   \mu^-$ candidates.

 Photon candidates   for $\chi_{c1}\to J/\psi \, \gamma$ and $\pi^0
\to \gamma \gamma$ decays are required to have an energy of at least
30  MeV in the barrel region ($|\cos\theta_{\gamma}|<0.71$) and at
least 50 MeV in the endcap region ($0.71<|\cos\theta_{\gamma}|<0.95$),
where $\theta_{\gamma}$ is the angle between the beam axis and the
candidate photon.  To select the  $\pi^0$ candidates for $B^0\to
J/\psi \,  \pi^0$ reconstruction, we   require the  normalized  $\pi^0
\to \gamma \gamma$ mass to be between $-5$ and $+4$. The average
$\gamma \gamma$ invariant mass resolution  for these $\pi^0$
candidates is   7~MeV$/c^2$.   We perform a fit constraining  the mass
of each $\pi^0$   candidate to the world average value~\cite{PDG}.

We reconstruct $\chi_{c1}$ in the  $\chi_{c1}\to J/\psi \, \gamma$
decay mode.  Most of the  photons in $\Upsilon(4S) \to B \overline  B$
events   come from $\pi^0$ decays.  We therefore do not use a photon
if it  can be paired with another photon to produce a $\pi^0$
candidate with the  normalized $\pi^0 \to \gamma \gamma$  mass between
$-4$ and $+3$. The resolution     in the $J/\psi \, \gamma$
invariant    mass  is 8~MeV$/c^2$.  We select the $\chi_{c1}$
candidates with   the normalized $\chi_{c1} \to J/\psi \, \gamma$
mass between $-4$  and $+3$ and     perform a fit constraining  the
mass of each $\chi_{c1}$   candidate to the world average
value~\cite{PDG}. 

The  $K^0_S \to \pi^+\pi^-$ candidates are selected   from  pairs of
tracks  forming well-measured  displaced vertices.   We refit the
daughter  pion   tracks taking into account the position of the
displaced vertex and constrain them   to  originate from   the
measured vertex.  The resolution     in the $\pi^+   \pi^-$  invariant
mass  is  4~MeV$/c^2$. We select the $K^0_S \to \pi^+\pi^-$ candidates
with  the normalized  $K^0_S \to \pi^+\pi^-$  mass between $-4$ and
$+4$ and  perform a fit constraining  the mass of each $K^0_S$
candidate to the world average value~\cite{PDG}.

In order to increase our $B^0 \to  J/\psi \, K^0_S$ sample, we also
reconstruct  $K^0_S \to \pi^0\pi^0$ decays. The average flight
distance for the $K^0_S$ from $B^0 \to  J/\psi \, K^0_S$ decay is
9~cm.  We  find the  $K^0_S$ decay vertex using only the calorimeter
information and the known position of the $e^+e^-$ interaction point.
The $K^0_S$ flight direction is calculated as the line passing through
the  $e^+e^-$ interaction point and the center of energy of the   four
photon showers in the calorimeter. The $K^0_S$ decay vertex is defined
as the point along  the $K^0_S$ flight direction  for which the
product $f[M(\gamma_1 \gamma_2)] \times f[M(\gamma_3 \gamma_4)]$ is
maximal.  In the above expression, $M(\gamma_1 \gamma_2)$ and
$M(\gamma_3 \gamma_4)$ are the diphoton invariant masses  recalculated
assuming a particular $K^0_S$ decay point and $f(M)$ is   the $\pi^0$
mass lineshape obtained from the simulation where we  use the known
$K^0_S$ decay vertex.  For simulated events, the $K^0_S$ flight
distance is found without  bias  with  a resolution of 5~cm.  The
uncertainty in the $K^0_S$ decay vertex position arising from the
$K^0_S$ direction approximation is much smaller than the resolution of
the flight distance. We select the $K^0_S$  candidates by requiring
the reconstructed $K^0_S$  decay length to be in the range from $-10$
to $+60$ cm.  After the $K^0_S$ decay vertex is found, we select the
$K^0_S \to \pi^0\pi^0$ candidates by requiring  $-15<M(\gamma
\gamma)-M_{\pi^0}<10$~MeV/$c^2$ for both photon pairs.  Then  we
perform a   kinematic fit simultaneously constraining $M(\gamma_1
\gamma_2)$ and $M(\gamma_3 \gamma_4)$ to the  world average value of
the $\pi^0$ mass~\cite{PDG}.  The resulting $K^0_S$ mass resolution is
12~MeV/$c^2$.  We select the $K^0_S \to \pi^0\pi^0$ candidates with
the normalized  $K^0_S \to \pi^0\pi^0$  mass between $-3$ and $+3$ and
perform a fit constraining  the mass of each $K^0_S$   candidate to
the world average value~\cite{PDG}.   The $K^0_S \to \pi^0\pi^0$
detection efficiency is determined  from simulation. The systematic
uncertainty associated with  this determination  can be reliably
estimated by comparing the $K^0_S \to \pi^0\pi^0$ and $K^0_S \to
\pi^+\pi^-$  yields for inclusive $K^0_S$ candidates in data and in
simulated events.

\begin{figure}[htb]
\centering \epsfxsize=165mm \epsfbox{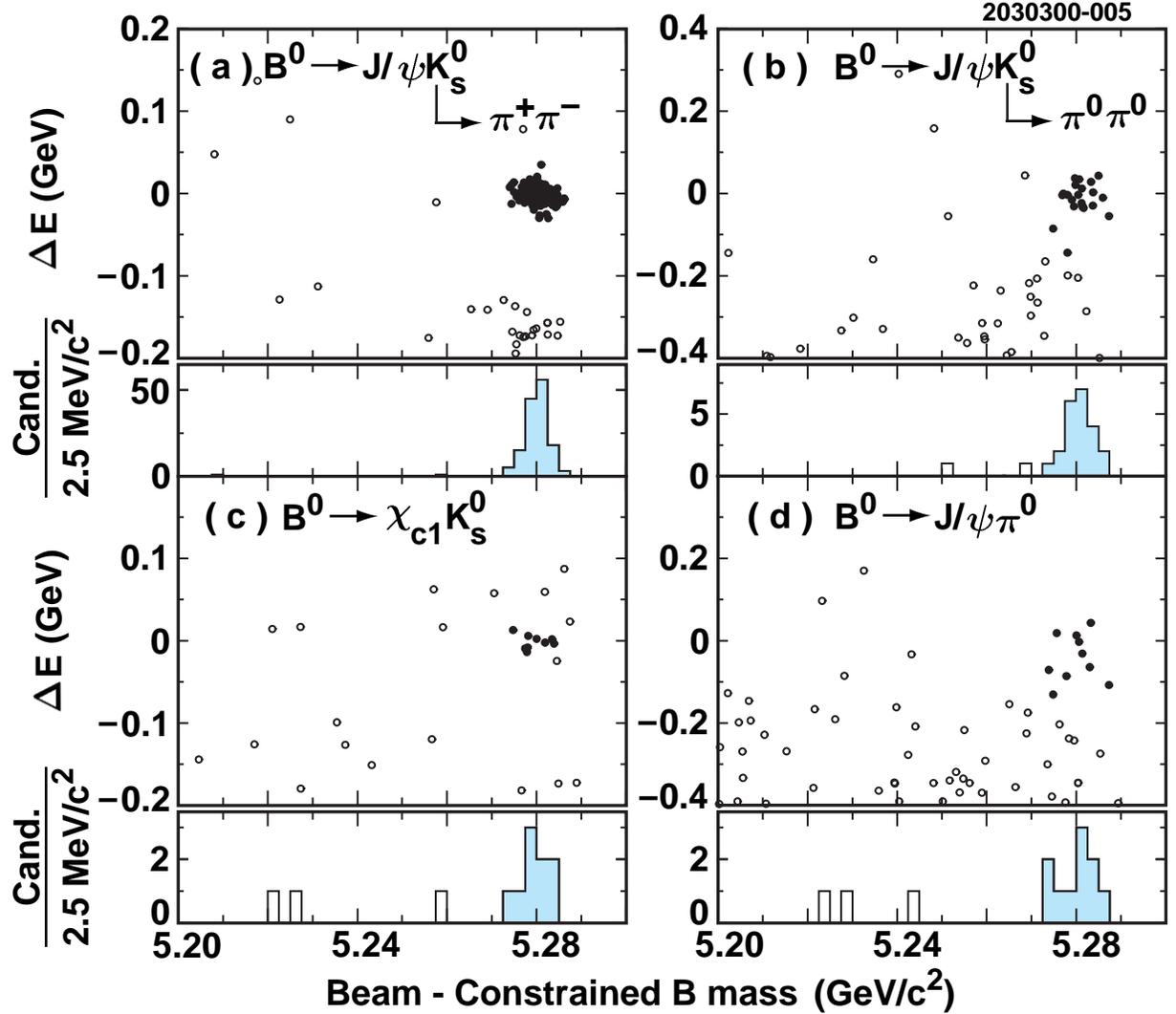}
\caption{  The  $\Delta E$  vs.  $M(B)$ distribution  for  (a) $B^0
\to J/\psi \, K^0_S$ with $K^0_S \to \pi^+ \pi^-$, (b) $B^0 \to J/\psi
\, K^0_S$ with  $K^0_S \to \pi^0 \pi^0$, (c) $B^0 \to \chi_{c1} \,
K^0_S$,  and (d) $B^0 \to J/\psi \, \pi^0$ candidates.   The signal
candidates, selected using normalized $\Delta E$  and $M(B)$
variables,  are  shown by  filled circles. Below each  $\Delta E$  vs.
$M(B)$  plot, we show  the projection on the $M(B)$ axis  with the
$\Delta E$ requirement  applied.  The shaded parts of the histograms
represent the   candidates  that pass the  $|M(B)-M_B|/\sigma(M)<3$
requirement.}
\label{fig:de_mb_all_data}
\end{figure}

 The $B^0$  candidates are  selected by means of two  observables. The
first  observable  is the difference between the energy of  the $B^0$
candidate and the  beam  energy,  $\Delta E  \equiv E(B^0) - E_{\rm
beam}$.  The average $\Delta E$ resolution for each decay mode  is
listed in Table~\ref{tab:results}.  We use the normalized  $\Delta E$
for  candidate selection and  require  $|\Delta E|/\sigma(\Delta E)<3$
for $B^0\to J/\psi \, K^0_S$ and $B^0\to \chi_{c1} \, K^0_S$
candidates with $K^0_S \to \pi^+\pi^-$.  To account for a low-side
$\Delta E$ tail arising from the energy leakage in the calorimeter, we
require  $-5<\Delta E/\sigma(\Delta E)<3$ for  $B^0\to J/\psi \,
K^0_S$ with  $K^0_S \to \pi^0\pi^0$ and $-4<\Delta E/\sigma(\Delta
E)<3$ for $B^0\to J/\psi \, \pi^0$ candidates.  The   second
observable   is  the     beam-constrained  $B$    mass,
$M(B)\equiv\sqrt{E^2_{\rm beam}-p^2(B)}$, where $p(B)$ is the
magnitude  of the $B^0$ candidate momentum. The  resolution in $M(B)$
is dominated by the beam energy spread for all the decay modes under
study  and varies from   2.7 to 3.0 MeV/$c^2$ depending on the mode.
We use the normalized  $M(B)$   for  candidate selection and require
$|M(B)-M_{B}|/\sigma(M)<3$, where $M_{B}$ is the nominal $B^0$ meson
mass.  The $\Delta E$  vs.  $M(B)$  distributions together with the
projections on the  $M(B)$ axis   are shown in
Fig.~\ref{fig:de_mb_all_data}.  The number of $B^0$ candidates
selected in each decay mode is listed in Table~\ref{tab:results}.

Backgrounds can be divided into two categories. The first category is
the  background from those exclusive $B$ decays that tend to produce a
peak in the signal region of the $M(B)$  distribution. We identify
these exclusive   $B$ decays  and estimate their contributions to
background       using  simulated events with the normalizations
determined from the known   branching fractions or from our data.  The
second category is the combinatorial background from $ B \overline B$
and continuum    non-$B \overline  B$   events. To estimate the
combinatorial background, we fit the $M(B)$ distribution in the region
from 5.1 to 5.3~GeV/$c^2$.  As a consistency check, we also estimate
the combinatorial background  using  high-statistics samples of
simulated $\Upsilon(4S) \to B \overline  B$ and non-$B \overline  B$
continuum  events  together with  the data collected below the $B
\overline B$ production threshold.  The total estimated backgrounds
are listed in Table~\ref{tab:results}.  Below we describe the
background estimation for each decay channel under study.   

{\it Background for $B^0\to J/\psi \, K^0_S$ with  $K^0_S \to
\pi^+\pi^-$.}  Only combinatorial background contributes, with the
total   background  estimated to be $0.3\pm0.2$ events.

{\it Background for $B^0\to J/\psi \, K^0_S$ with  $K^0_S \to
\pi^0\pi^0$.}  The combinatorial background  is estimated to be
$0.5\pm0.2$ events. The  other background source is  $B\to J/\psi \,
K^*$~\cite{K-pi-notation}, with $K^*\to K\pi^0$ or  $K^*\to K^0_S\pi$
with $K^0_S \to \pi^0\pi^0$. The background from these decays is
estimated to be $0.6\pm0.2$ events.

{\it Background for $B^0\to \chi_{c1} \, K^0_S$.}  The combinatorial
background is estimated to be $0.5\pm0.3$ events.  We estimate the
background from $B$ decays to  the $J/\psi \,K^0_S\pi$ final  state
from the samples of simulated events, with the  normalizations
obtained from  the fits to the $M(K\pi)$ distributions for $B^+ \to
J/\psi \,K^0_S \pi^+$ and $B^0 \to J/\psi \,K^- \pi^+$ candidates in
data.  The background from $B \to J/\psi \,K^0_S\pi$  is estimated to
be $0.41\pm0.07$ events, and is dominated by $B\to J/\psi \, K^*$
decays with  $K^*\to K^0_S \pi$. We find no evidence for $B\to
\chi_{c2} \, K$  production and estimate  the background from  $B^0
\to \chi_{c2} \, K^0_S$  to be  $0.01\pm0.01$ events.
  
{\it Background for $B^0 \to J/\psi \, \pi^0$.}  The combinatorial
background is estimated to be $0.4^{+0.5}_{-0.3}$ events.  The $\Delta
E$ resolution is good enough to  render negligible the background from
any of the Cabibbo-allowed $B \to J/\psi K \pi^0 (X)$ decays, where at
least a kaon mass   is missing from the energy sum.  The background
from $B^0\to J/\psi \, K^0_S$ decays with  $K^0_S \to \pi^0\pi^0$   is
estimated to be $0.38\pm0.05$ events. We estimate the background from
$B$ decays to  the $J/\psi \,\pi \pi^0$ final  state from the samples
of simulated events, with the  normalizations obtained from  the fits
to the $M(\pi \pi)$ distributions for $B^+ \to J/\psi \, \pi^+ \pi^0$
and $B^0 \to J/\psi \,\pi^0 \pi^0$ candidates in data.  The background
from $B \to J/\psi \,\pi^0 \pi$   is estimated to be $0.2\pm0.2$
events, and is dominated by $B^+ \to J/\psi \, \rho^+$  decays.  

We      use  the Feldman-Cousins approach ~\cite{Feldman:1998qc} to
assign the 68\%  C.L. intervals  for the signal mean for the three
low-statistics  decay modes ($B^0 \to \chi_{c1} \, K^0_S$, $B^0 \to
J/\psi \, \pi^0$, and  $B^0 \to J/\psi \, K^0_S$ with $K^0_S \to \pi^0
\pi^0$).  We assume ${\cal B}(\Upsilon(4S) \to B^0  {\overline
B^0})={\cal B}(\Upsilon(4S) \to B^+   B^-)$ for all branching
fractions in this Article.  We use the following branching fractions
for  the secondary decays: ${\cal B}(J/\psi \to \ell^+
\ell^-)=(5.894\pm0.086)\%$~\cite{Bai:1998di},  ${\cal B}(\chi_{c1}\to
J/\psi \gamma)=(27.3\pm1.6)\%$~\cite{PDG}, ${\cal B}(K^0_S \to \pi^+
\pi^-)=(68.61\pm0.28)\%$~\cite{PDG}, and ${\cal B}(K^0_S \to \pi^0
\pi^0)=(31.39\pm0.28)\%$~\cite{PDG}.  The reconstruction efficiencies
are determined from  simulation.  The resulting branching fractions
are listed in  Table~\ref{tab:results}.  Combining the results for the
two $K^{0}_S$ modes used in $B^0\to J/\psi \, K^0_S$ reconstruction
and taking into account correlated systematic uncertainties, we obtain
${\cal B}(B^0 \to J/\psi \, K^0)  = (9.5 \pm 0.8 \pm
0.6)\times10^{-4}$. The measurements of  ${\cal B}(B^0 \to J/\psi \,
K^0)$, ${\cal B}(B^0 \to \chi_{c1} \, K^0)$, and  ${\cal B}(B^0 \to
J/\psi \, \pi^0)$ reported in this Article   supersede the previous
CLEO results~\cite{CLEO-previous}.

The  systematic uncertainties  in the branching fraction measurements
include   contributions from the  uncertainty in the number of $B
\overline B$  pairs (2\%),   tracking efficiencies (1\% per charged
track), photon   detection efficiency (2.5\%),  lepton  detection
efficiency (3\% per lepton),  $K^0_S\to\pi^+\pi^-$ finding efficiency
(2\%), $K^0_S\to\pi^0\pi^0$ finding efficiency (5\%), background
subtraction ($0.01-5.5\%$, see Table~\ref{tab:results}), statistics
of the simulated  event  samples ($0.6-1.0\%$), and  the
uncertainties   on the  branching fractions of secondary decays (see
Table~\ref{tab:results}).  

In summary, we have studied three $B^0$ decay modes useful for the
 measurement of $\sin2\beta$.  We report the first observation and
 measure branching  fractions of  the  $B^0 \to \chi_{c1} \, K^0$ and
 $B^0 \to J/\psi \, \pi^{0}$ decays.  We describe  a $K^{0}_S \to
 \pi^0 \pi^0$ detection technique  and its application to the
 reconstruction   of the decay $B^0 \to J/\psi \, K^{0}_S$.  We
 measure the  branching fraction  for  $B^0 \to J/\psi \, K^0$ decays
 with  $K^0_S$ mesons reconstructed in both $\pi^+\pi^-$ and
 $\pi^0\pi^0$ decay modes. 

We gratefully acknowledge the effort of the CESR staff in providing us
with excellent luminosity and running conditions.  This work was
supported by  the National Science Foundation, the U.S. Department of
Energy, the Research Corporation, the Natural Sciences and Engineering
Research Council of Canada,  the A.P. Sloan Foundation,  the Swiss
National Science Foundation,  the Texas Advanced Research Program, and
the Alexander von Humboldt Stiftung.

\begin{table}[htb]
\center
\caption{\small Number of signal candidates, estimated background,
average $\Delta E$ resolution,  product of secondary branching
fractions (${\cal B}_s$), detection  efficiency, and  measured
branching fraction. Row 1 contains the combined value of ${\cal B}(B^0
\to J/\psi \, K^0)$,   rows 2 and 3 contain  the individual results
for the two $K^0_S$ decay modes.}\label{tab:results}
\begin{tabular}{lllllll} 
  Decay           & Signal    & Total  &  $\sigma(\Delta E)$ &
~~~~${\cal B}_s$    & Efficiency    & Branching   \\  mode           &
candidates     &      background  &         (MeV)  & ~~~$(\%)$ &
~~~$(\%)$  & fraction  $(\times10^{-4})$  \\ \hline $B^0 \to J/\psi \,
K^0$      &           &         &  &         &              &
$9.5\pm0.8\pm0.6$   \\  ~~~$K^0_S \to \pi^+ \pi^-$      &   142      &
$0.3\pm0.2$ &   11 &   $4.04\pm0.06$    & $37.0\pm2.3$     & $9.8 \pm
0.8 \pm 0.7$                 \\ ~~~$K^0_S \to \pi^0 \pi^0$     &   ~22
&  $1.1\pm0.3$     &      25\tablenotemark[1]   & $1.85\pm0.03$     &
$13.9\pm1.1$\tablenotemark[2]     & $8.4^{+2.1}_{-1.9} \pm 0.7$
\\   $B^0 \to \chi_{c1}\,  K^0$   &   ~~9      & $0.9\pm0.3$  &  10  &
$1.10\pm0.07$     & $19.2\pm1.3$     & $3.9^{+1.9}_{-1.3}\pm 0.4$
\\ $B^0 \to J/\psi \, \pi^0$       &   ~10      & $1.0\pm0.5$  &
28\tablenotemark[1]  & $11.8\pm0.2$     &
$31.4\pm2.2$\tablenotemark[2]     & $0.25^{+0.11}_{-0.09}\pm 0.02$
\\               
\end{tabular}
\flushleft \tablenotetext[1]{The $\Delta E$ distribution has a
low-side tail due to the energy leakage in the calorimeter.}
\tablenotetext[2]{Includes the  loss of efficiency due to $\pi^0\to
e^+ e^-\gamma$ decays.}
\end{table}

\begin{thebibliography}{99}
\bibitem{K-pi-notation} We refer to $K^*(892)$ as $K^*$. Unless
otherwise noted, $K$ is either  $K^+$ or $K^0$; similarly $\pi$ is
either  $\pi^+$ or $\pi^0$.  Charge conjugation is implied.
\end{thebibliography}
\end{document}